\documentclass[aps,prl,twocolumn,showpacs,groupedaddress]{revtex4}
\usepackage{graphicx}

\begin{document}

\title{Berry phase in a composite system}
\author{X. X. Yi$^1$, L. C. Wang$^2$, and T. Y. Zheng$^2$}
\affiliation{$^1$Department of physics, Dalian University of
Technology, Dalian 116024, China\\
$^2$Institute of Theoretical Physics, Northeast Normal University,
Changchun 130024, China}

\date{\today}

\begin{abstract}
The Berry phase in a composite system with only one  subsystem
being driven has been studied in this Letter. We choose two
spin-$\frac 1 2 $ systems with spin-spin couplings as the
composite system, one of the subsystems is driven by a
time-dependent magnetic field. We show how the Berry phases depend
on the coupling between the two subsystems, and what is the
relation between these Berry phases of the whole system and those
of the subsystems.

\end{abstract}

 \maketitle

Geometric phases in quantum theory attracted great interest since
Berry  \cite{berry} shown that the state of a quantum system
acquires a purely geometric feature in addition to the usual
dynamical phase when it is varied slowly and eventually brought
back to its initial form. The Berry phase has been extensively
studied \cite{shapere, thouless, sun} and generalized in various
directions \cite{aharonov, sjoqvist1, ericsson, carollo, fuentes},
for example, geometric phases for mixed states \cite{sjoqvist1},
for open systems\cite{carollo}, and with a quantized field driving
\cite{fuentes}. In a recent paper \cite{sjoqvist2}, Sj\"oqvist
calculated the geometric phase for a pair of entangled spins in a
{\it time-independent} uniform magnetic field. This is an
interesting development in holonomic quantum computer and in
addition it shown how the prior entanglement in an initial state
modify the Berry phase. This study was generalized \cite{tong1} to
the case of spin pairs in a {\it rotating } magnetic field, the
result shown  that the geometric phase of the whole entangled
bipartite system can be decomposed into a sum of  geometric phases
of the two subsystems, provided the evolution is cyclic.

Roughly speaking, entanglement may be created only via
interactions or jointed measurements, thus how subsystem-subsystem
interaction change the Berry phases of a composite system and
those of the subsystems is of interest, from both sides of
experimental and theoretical viewpoint. On the other hand, the
Berry phase has very interesting applications, such as the
implementation of quantum computation by geometric
means\cite{jones1, ekert,falci,wang}. All systems for this purpose
are composite, i.e., it at least consists of two subsystems with a
direct coupling or being coupled through a third party. This again
gives rise to questions of how the couplings among the subsystems
changes the Berry phase of the composite system and what is the
relation between  these Berry phases of the composite system and
those of the two subsystems.

In this Letter, we investigate the behavior of the Berry phase of
two spin-$\frac 1 2 $ systems with spin-spin couplings, one of the
spin-$\frac 1 2 $ is driven by a time-dependent magnetic field
precessing around the z-axis. We calculate and analyze the effect
of spin-spin coupling on the Berry phase of the composite system
and those of the subsystems.  As you will see, the Berry phase of
the composite system is just a sum over those of the subsystems.
This result is completely general, although we use spin half as an
example to demonstrate the feature of the Berry phase.  The
Hamiltonian describing a system consisting of two interacting
spin-$\frac 1 2 $ particles in the presence of an external
magnetic field takes the form,
\begin{equation}
H=\frac1 2 \alpha \vec{\sigma}_1 \cdot \vec{B}(t)+
J(\sigma_1^+\sigma_2^++h.c.),
\end{equation}
where $\vec{\sigma}_j=(\sigma^x_j,\sigma^y_j, \sigma^z_j)$,
$\sigma^i_j$ are the pauli operators  for subsystem $ j (j=1,2)$
and $\sigma_j^+=(1/2)(\sigma_j^x+i\sigma_j^y).$ We will choose
$\vec{B}(t)=B_0 \hat{n}(t)$ with the unit vector
$\hat{n}=(\sin\theta\cos\phi,\sin\theta\sin\phi,\cos\theta)$ and
have assumed  that only the subsystem 1 is driven by external
fields. The classical field $\vec{B}(t)$ acts as an external
control parameter, as its direction and magnitude can be
experimentally changed. $J$ stands for the constant of coupling
between the two spin-$\frac 1 2 $.  This coupling is not a typical
spin-spin coupling, but rather a toy model describing a double
spin flip; nevertheless,  the presentation in this Letter can be
generalized to the system of nuclear magnetic resonance(NMR) in
which quantum computation is implemented by geometric means
\cite{jones1}, furthermore  the observation of geometric phase for
such a system  is feasible by the current  technology\cite{du}.

In a space spanned by $\{ |eg\rangle,|ee\rangle,|gg\rangle,
|ge\rangle\}$ and in units of $\frac 1 2 \alpha B_0$, the
Hamiltonian Eq.(1) can be written as

\begin{equation}
H=\left( \matrix{ \cos\theta & 0 & \sin\theta e^{-i\phi} &0 \cr
 0 & \cos\theta & g & \sin\theta e^{-i\phi}\cr
 \sin\theta e^{i\phi} &g & -\cos\theta &0 \cr
 0 & \sin\theta e^{i\phi} & 0 & -\cos\theta \cr } \right),
\end{equation}
with $g=\frac{2J}{\alpha B_0}$ a rescaled  coupling constant.
Keeping $\theta$ constant and  changing $\phi$ slowly from $0$ to
$\phi(T)=2\pi$ the Berry phase generated after the system
undergoing  an adiabatic and cyclic evolution starting with an
initial state  $|\Psi_j(t=0)\rangle$ may be calculated as follows:
\begin{equation}
\gamma_j=i\int_0^T dt\langle\Psi_j|\dot{\Psi_j}\rangle,
\label{berryp}
\end{equation}
where $|\Psi_j\rangle$ $(j=1,2,3,4)$ are the instantaneous
eigenstates of the Hamiltonian Eq.(1) and have the following form
\begin{eqnarray}
|\Psi_j\rangle&=&\frac{1}{\sqrt{M_j}}[a_j(\phi,\theta,g)|eg\rangle+
b_j(\phi,\theta,g)|ee\rangle\nonumber\\
&+&c_j(\phi,\theta,g)|gg\rangle+d_j(\phi,\theta, g)|ge\rangle],
\label{eigenv1}
\end{eqnarray}
with
\begin{eqnarray}
a_j(\phi,\theta,g)&=& \sin\theta e^{-i\phi},
c_j(\phi,\theta,g)=E_j-\cos\theta,\nonumber\\
d_j(\phi,\theta,g)&=&\frac{g(\cos\theta-E_j)}{\sin\theta-(\cos\theta-E_j)\frac{\cos\theta+E_j}{\sin\theta}}
e^{i\phi},\nonumber\\
b_j(\phi, \theta,
g)&=&-\frac{\cos\theta+E_j}{\sin\theta}e^{-i\phi}
d_j(\phi,\theta,g),\nonumber\\
M_j&=&|a_j|^2+|b_j|^2+|c_j|^2+|d_j|^2,\label{eigenf}
\end{eqnarray}
and $E_j$ that denote the instantaneous eigenvalues of the
Hamiltonian Eq.(1) can be calculated by solving
$$(\cos^2\theta-E_j^2)^2+(2\sin^2\theta+g^2)(\cos^2\theta-E_j^2)+\sin^4\theta=0.$$
In the simplest case, where the coupling constant $g=0$, the
eigenvalues $E_{\pm}=\pm 1$, the corresponding  eigenstates follow
from Eq.(\ref{eigenf}) that $a_+=\sin\theta e^{-i\phi},$ $
c_+=1-\cos\theta,$ $ b_+=d_+=0$ and $a_-=\sin\theta e^{-i\phi},$ $
c_-=-1-\cos\theta,$ $ b_-=d_-=0.$ These give rise to the well
known Berry phase $\gamma_+=\pi(1+\cos\theta)$ and
$\gamma_-=\pi(1-\cos\theta)$. This result  is easy to understand,
 the subsystem 2 that evolves freely has no effects on any
behaviors of the  subsystem 1 as long as the whole system is
initially prepared in a separable state. Hence, the Berry phase of
the composite system is exactly that of the subsystem 1, while the
subsystem 2 acquires no geometric phase. For a noncyclic and
non-adiabatical process, the author \cite{sjoqvist2, tong2} draw
out the same results for geometric phases. We will now turn to
study the effect of the coupling between subsystems 1 and 2 on the
Berry phase of the whole system, first of all we write down the
four eigenvalues of the Hamiltonian as
\begin{eqnarray}
E_1&=&\sqrt{1+\frac {g^2}{ 2} +\frac g 2
\sqrt{g^2+4\sin^2\theta}}=-E_2,
\nonumber\\
E_3&=&\sqrt{1+\frac {g^2}{2} -\frac g 2
\sqrt{g^2+4\sin^2\theta}}=-E_4. \label{eigenv}
\end{eqnarray}
Substituting these eigenvalues into Eqs (\ref{eigenf}) and
(\ref{berryp}), we can get respective Berry phases. The dependance
of the Berry phase on the coupling constant as well as on the
azimuthal angle $\theta$ was illustrated in Figures from 1 to 5.
Figures from 1 to 4 are for the Berry phases with varying coupling
constant $g$ and azimuthal angle $\theta$, whereas figure 5 shows
the dependance of the Berry phase on the coupling constant with a
specific azimuthal angle $\theta=\frac{\pi}{4}$. The common
feature of these figures is that  with the rescaled coupling
constant $g\rightarrow \infty$,  all Berry phases
$\gamma_i\rightarrow 0$ ( All phases are defined modulo $2\pi$
throughout this paper). This limit corresponds to the case when
the first term in Hamiltonian Eq.(1) can be ignored. Physically,
the spin-spin coupling may modify the azimuthal angle to an
effective one  with which the system precessing around the z-axis.
 The spin-spin couplings describe a jointed spin flip
of the subsystems,  the coupling constant then characterizes the
flip frequency, consequently,  the effective azimuthal angle
should be an average over all possible azimuthal angles which
would take positive and negative values with equal probabilities
in the limit $g\rightarrow \infty $.

\begin{figure}
\includegraphics*[width=0.95\columnwidth,
height=0.6\columnwidth]{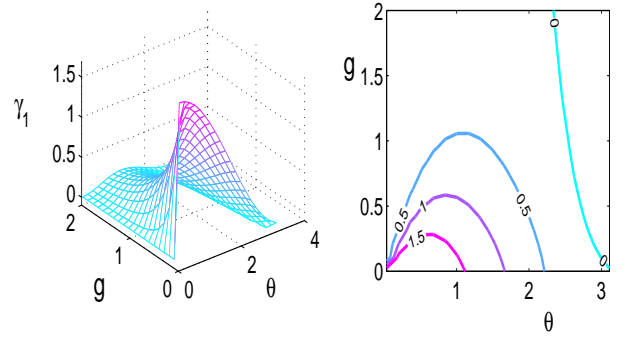} \caption{(Color online) Berry
phase corresponding to the instantaneous eigenstate
$|\Psi_1\rangle$ versus the rescaled coupling constant $g$ and the
azimuthal angle $\theta$[Arc]. The plot was presented in units of
$\pi$ for the Berry phase.  The right panel is a contour plot for
the left one. } \label{fig1}
\end{figure}
From Fig.1 we see that the Berry phase is a monotonic function of
the rescaled coupling constant, while it is maximized for
intermediate values of the azimuthal angle $\theta$. The berry
phase for the eigenstate 2 has a similar appearance  as Fig.2
shows.
\begin{figure}
\includegraphics*[width=0.95\columnwidth,
height=0.6\columnwidth]{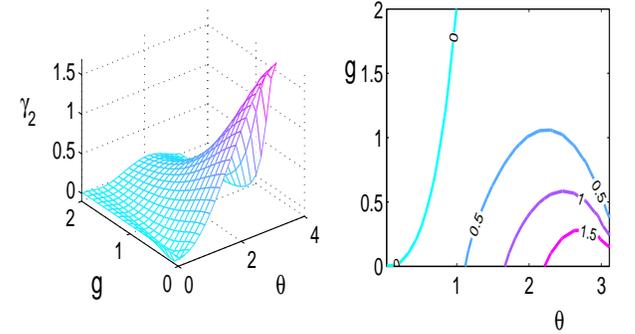} \caption{ (Color online) The
same as Figure 1, but for instantaneous eigenstate
$|\Psi_2\rangle$ } \label{fig1}
\end{figure}
It is worth  noting that $\gamma_1(\theta)=\gamma_2(\pi-\theta)$,
this can be easily found by comparing the contour plots presented
in Fig. 1 and 2. This symmetry originate from the Hamiltonian and
it  is clear that the eigenstate $|\Psi_1\rangle (|\Psi_3\rangle)$
is alternated with $|\Psi_2\rangle (|\Psi_4\rangle)$ when $\theta\
\leftrightarrow (\pi- \theta)$, this leads to the symmetry in the
Berry phase.  The contour plots presented in Fig.3 and 4 show the
same symmetry indeed.

\begin{figure}
\includegraphics*[width=0.95\columnwidth,
height=0.6\columnwidth]{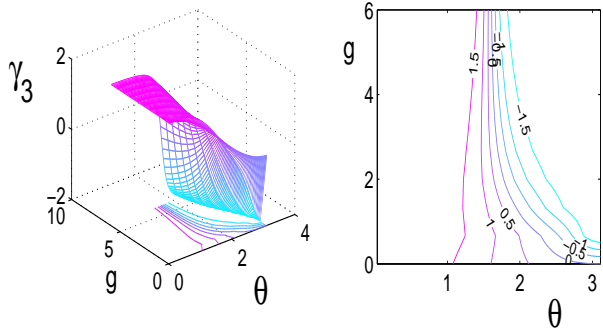} \caption{(Color online) Berry
phase (in units of $\pi$)  corresponding to the instantaneous
eigenstate $|\Psi_3\rangle$ versus the rescaled coupling constant
$g$ and the azimuthal angle $\theta$[Arc]. The right panel is a
contour plot for the left one. } \label{fig1}
\end{figure}

\begin{figure}
\includegraphics*[width=0.95\columnwidth,
height=0.6\columnwidth]{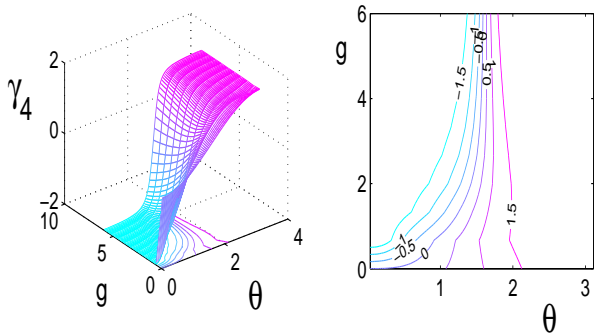} \caption{(Color online) The same
as Figure 3, but for the instantaneous eigenstate $|\Psi_4\rangle$
} \label{fig1}
\end{figure}

 Figure 5 shows the
results of Berry phases corresponding to the four eigenstates
Eq.(4) with a specific azimuthal angle
$\theta=\theta_0=\frac{\pi}{4}$. With $g\rightarrow 0$ the Berry
phases approach two values( in units of $\pi$) of
$\gamma_{\pm}=(1\pm \cos\theta_0)\simeq 1\pm 0.707$ as expected,
whereas they approach zero with $g\rightarrow \infty$.

\begin{figure}
\includegraphics*[width=0.95\columnwidth,
height=0.6\columnwidth]{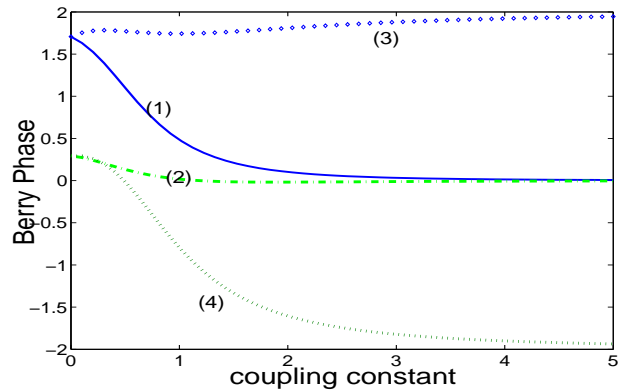} \caption{(Color online)
 The Berry phase {\it vs.} the rescaled  coupling constant $g$ with a specific azimuthal
 angle $\theta=\frac{\pi}{4}$. The indices on the line indicate the eigenstate
 by which we got the Berry phase.} \label{fig1}
\end{figure}
Now we are in a position to study the Berry phase of the
subsystem, and to show what is the relation between these phases.
Generally speaking a state of  subsystem is no longer a pure one,
so we have to adopt the definition of geometric phase for a mixed
state \cite{sjoqvist1}, that is  $\phi_g=\arg \mbox{Tr}(\rho_0
U(t))$ with $\rho_0$ denoting the initial density matrix and
$U(t)$ the transport operator which should fulfill the parallel
transport evolution condition. This definition is available when
the system from which we want to get geometric phases undergoes an
unitary evolution. For the subsystems with non-zero couplings,
however, the evolution of each subsystem is not unitary in
general. So, here we borrow the idea presented in \cite{ericsson}
to define the Berry phase for a mixed state. A non-unitary
evolution of a quantal state may be conveniently modelled by
attaching an ancilla to the system, in our case the ancilla can
always be taken to be the other spin-$\frac 1 2 $ system. The
geometric phase corresponding to this non-unitary evolution is
then defined as the geometric phase of the whole system (system+
ancilla) that evolves unitarily. For an adiabatic cyclic
evolution, this gives rise to a definition of Berry phase for a
mixed state $\rho(t)=\sum_jp_j|E_j(t)\rangle\langle E_j(t)|,$
\begin{equation}
\gamma=\sum_jp_j\gamma_j,
\end{equation}
where $\gamma_j=i\int_0^T dt \langle E_j(t)| \dot{E}_j(t)\rangle$.
The Berry phase Eq.(7) for a mixed state is just an average of the
individual Berry phases, weighted by their eigenvalues $p_j$.  To
be sure, what we have is consistent with known results, we check
that this expression reduces to the standard Berry phase
$\gamma=i\int_0^T dt\langle \psi(t)|\dot{\psi}(t)\rangle$ for a
pure state $\rho(t)=|\psi(t)\rangle\langle \psi(t)|$. In our case,
we have four density matrices of mixed state for each subsystem,
they correspond to the four instantaneous eigenstates of the
Hamiltonian, respectively. For example,
$\rho_1^j(t)=\mbox{Tr}_2|\Psi_j(t)\rangle\langle \Psi_j(t)|$
represents the $j$-th density matrix for subsystem 1 among the
four density matrices, where $\mbox{Tr}_2$ denotes a trace over
subsystem 2. The Berry phase corresponding to this state
$\rho_1^j(t)$  is then given by Eq.(7). Actually, the definition
Eq.(7) can be derived by the idea of  the so-called purifications
as follows. We may construct a pure state
$$
|\Phi(t)\rangle=\sum_j\sqrt{p_j}|E_j(t)\rangle_1\otimes|j\rangle_a$$
for subsystem 1 + ancilla (for subsystem 2+ ancilla, in the same
manner) such that $$\mbox{Tr}_a|\Psi(t)\rangle\langle
\Psi(t)|=\sum_jp_j|E_j(t)\rangle\langle E_j(t)|=\rho_1(t),$$ where
$\mbox{Tr}_a$ denotes a trace over the ancilla and
$|E_j(t)\rangle$ represent instantaneous eigenstates of
$\rho_1(t)$. Since the states of the ancilla remain unchanged
during the evolution, the Berry phase of the subsystem 1 is then
 the Berry phase of the compound
(subsystem+ancilla), this yields the definition Eq.(7).
\begin{figure}
\includegraphics*[width=0.95\columnwidth,
height=0.6\columnwidth]{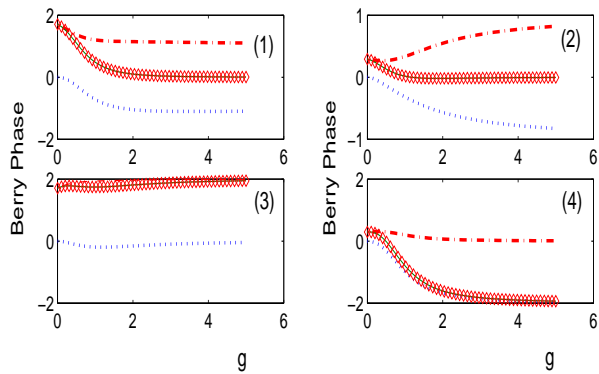} \caption{ (Color online) The
Berry phase for the composite system(diamond) and for subsystem
1(dash-dot) and subsystem 2 (dotted). A sum of the subsystem's
Berry phase is also presented in the figure(solid line, overlapped
by the diamond line). The azimuthal angle $\theta=\frac{\pi}{4}$
was chosen for this plot.} \label{fig1}
\end{figure}
The Berry Phase of the composite system and those of the two
subsystems are illustrated in Fig.6, a sum of the subsystem's
Berry phase is also shown. There is an evidence that the Berry
phase of the composite system can be decomposed into a sum of the
subsystem's Berry phases, it reveals the relation between
geometric  phases of entangled bipartite systems and those of
their subsystems. We can prove this point indeed by expanding the
instantaneous eigenstate of the composite system via Schmidt
decomposition
\begin{equation}
|\Psi\rangle=\sum_i\sqrt{p_i}|e_i(t)\rangle_1\otimes|E_i(t)\rangle_2,
\label{sch}
\end{equation}
where  $|\Psi\rangle$ denotes one of the instantaneous eigenstates
Eq.~(\ref{eigenv1}). This expansion yields the reduced density
operator $\rho_1(t)=\sum_i p_i|e_i(t)\rangle\langle e_i(t)|$ and
$\rho_2(t)=\sum_i p_i|E_i(t)\rangle\langle E_i(t)|$ for subsystems
$1$ and $2$, respectively. By the definition Eq.~ (\ref{berryp}),
the Berry phase corresponding to $|\Psi\rangle$ follows,
\begin{equation}
\gamma=i\int_0^T\sum_jp_j\langle e_j(t)|\dot{e}_j(t)\rangle
dt+i\int_0^T\sum_jp_j\langle E_j(t)|\dot{E}_j(t)\rangle dt,
\end{equation}
i.e., the Berry phase of the composite system adds up to be that
of the composite system. This additivity holds mathematically when
the Schmidt decomposition is available with {\it time-independent}
coefficients $p_i$. Physically, time-independent coefficients
$p_i$ indicate no population transfer among the eigenstates of the
reduced density matrix of subsystem (this is what we called
adiabaticity for the subsystem in this paper). Here the result
that the Berry phase of the subsystem adds up to be that for the
composite system remains unchanged for all two-subsystem compound
when both the compound and the subsystems undergo  a cyclic
adiabatic evolution.

The observation of this prediction with NMR experiment is within
the touch of current technology \cite{du}. For instance, we can
use Carbon-13 labelled chloroform in $d_6$ acetone as the sample,
in which the single $^{13}C$ nucleus and the $^1H$ nucleus play
the role of the two spin-$\frac 1 2 $. The  constant of spin-spin
coupling $J\sigma_1^z\sigma_2^z$  in this case is  $J\simeq
(2\pi)214.5 \mbox{Hz}$, and we may control the rescaled coupling
constant $g$ by changing the magnitude of the external magnetic
field. We would like to address that the interaction between the
two spin-$\frac 1 2 $ in our model is not a typical spin-spin
coupling as that in NMR, but rather a toy model describing a
double spin flip. So, we have to make a mapping when we employ the
presentation in NMR system and when all subsystem are driven by
the classical field.  Finally, we want to discuss the problem of
adiabaticity. Our study is based on an adiabatic cyclic evolution
of the composite system. For any subsystem, however, the
conditions of adiabaticity are not fulfilled in general, in this
sense this is not a Berry phase but a geometric phase for a
subsystem when the composite system itself is subject to an
adiabatic evolution. But this is not the case in the Letter, it is
easily to check that the eigenvalues of the reduced density matrix
$\rho_1^j(t)=\mbox{Tr}_2(|\Psi_j(t)\rangle\langle\Psi_j(t)|$ (for
any j) are independent of time, this indicate the population on
the eigenstate of $\rho_1^j(t)$ remain unchanged \cite
{note1}while the composite system follows an adiabatic evolution.

To sum up, we have theoretically investigated the Berry phase of a
composite system and that of their subsystems. The Berry phase for
a mixed state to our best knowledge is a concept new project. The
relation between those phases is also presented and discussed,
these results provide us a new way to control the Berry phase, we
thus might find some applications in quantum computation. We are
investigating possible applications of this effects and its
connections to other quantum effects in different systems.

\ \ \\
XXY acknowledges enlightening discussions with Dr Erik
Sj\"{o}qvist and Dr Jiangfeng Du.  This work is supported
by EYTP of M.O.E, and NSF of China Project No. 10305002.\\

\end{document}